\documentstyle[12pt,aaspp4]{article}

\lefthead{Ohyama \& Taniguchi}
\righthead{MOLECULAR-GAS SUPERWIND FROM THE FACE-ON WOLF-RAYET GALAXY MRK 1259}

\begin{document}

\title{MOLECULAR-GAS SUPERWIND FROM THE FACE-ON WOLF-RAYET GALAXY MRK 1259}

\author{Youichi Ohyama, \& Yoshiaki Taniguchi}

\affil{Astronomical Institute, Tohoku University, Aoba, Sendai 980-8578, Japan}
%\begin{center}
%{\it To appear in the Astrophysical Journal, Letters}
%\end{center}

\begin{abstract}

We report results of $^{12}$CO ($J=1$-0) mapping observations of the
Wolf-Rayet starburst galaxy Mrk 1259 which has optical evidence for
the superwind seen from a nearly pole-on view.
The CO emission is detected in the central 4 kpc region.
The nuclear CO spectrum shows a blue-shifted
($\Delta V \simeq -27$ km s$^{-1}$) broad
(FWHM $\simeq$ 114 km s$^{-1}$) component as well as the
narrow one (FWHM $\simeq 68$ km s$^{-1}$).
The off-nuclear CO spectra also show the single-peaked broad component
(FWHM $\simeq$ 100 km s$^{-1}$).
The single-peaked CO profiles of both the nuclear and off-nuclear
regions may be explained if we introduce a CO gas disk with
a velocity dispersion of $\sim 100$ km s$^{-1}$.
If this gas disk would be extended up to a few kpc in radius,
we may explain the wide line widths of the off-nuclear CO emission.
Alternatively, we may attribute the off-nuclear CO emission to the gas
associated with the superwind.
However, if all the CO gas moves along the biconical surface of the superwind,
the CO spectra would show double-peaked profiles.
Hence, the single-peaked CO profiles of the off-nuclear regions may be
explained by an idea that the morphology and/or velocity field  of
the molecular-gas superwind are more complex as suggested by hydrodynamical
simulations.

\end{abstract}

%------------------------------------------------------------------------------------


%-----------------------------------------------------------------------------------

\section{INTRODUCTION}

In starburst galaxies, a large number of massive stars
(e.g., $\sim 10^{4-5}$) are formed within a short duration
(Weedman et al. 1981; Balzano 1983; Taniguchi et al. 1988).
Therefore, a burst of supernova explosions occurs inevitably
$\sim 10^7$ years after the onset of the starburst.
Since these numerous supernovae release a huge amount of kinetic
energy into the circumnuclear gas,
the circumnuclear gas is thermalized and then blow out into
the direction perpendicular to the galactic disk as a ``superwind''
(Tomisaka \& Ikeuchi 1988; Heckman, Armus, \& Miley 1990; Suchkov et al. 1994).
A bubble of the ionized gas sweeps up the circumnuclear molecular gas,
leading to the formation of molecular-gas superwind as well as
the ionized-gas one (Tomisaka \& Ikeuchi 1988; Suchkov et al. 1994).
Thus, in order to understand the whole physical processes of superwinds,
it is important to investigate the nature of molecular-gas superwinds
(e.g., Nakai et al. 1987; Aalto et al. 1994; Irwin \& Sofue 1996).

In this {\it Letter}, we present new evidence for the molecular-gas superwind
from the Wolf-Rayet starburst galaxy Mrk 1259, which shows
the optical evidence for the superwind viewed from a nearly pole-on view
(Ohyama, Taniguchi, \& Terlevich 1997; hereafter Paper I).
Mrk 1259 is a peculiar S0 galaxy (de Vaucouleurs et al. 1991; hereafter RC3)
at a distance of 26.64 Mpc
\footnote{Paper I adopted a distance toward Mrk 1259, $D = 33.5$ Mpc.
However, $V_{\rm 3K}$ was misused instead of $V_{\rm GSR}$ in this estimate.
In this {\it Letter}, using $V_{\rm GSR} = 1998$ km s$^{-1}$ (RC3),
with a Hubble constant $H_0$ = 75 km s$^{-1}$ Mpc $^{-1}$,
we adopt a distance $D$ = 26.64 Mpc.
Therefore, the HeII$\lambda$4686 luminosity, the number of late WR (WRL) stars,
the size of the superwind, and the average velocity of the superwind
in Paper I should be read as
$L$(HeII) = 7.0$\times 10^{39}$ erg s$^{-1}$,
$N$(WRL) $\simeq$ 4100,
$r$(superwind) $\simeq 3.3$ kpc,
and the average wind velocity $\simeq$ 565 km s$^{-1}$, respectively.}.
The logarithmic major-to-minor diameter ratio,
log $R_{\rm 25}=0.10\pm 0.08$ (RC3),
gives a nominal inclination angle, $i=37\fdg 4^{+11.2}_{-20.1}$, and
the galaxy appears to be elongated along the EW direction.
If this elongation were attributed to the inclination, we would observe
the rotational motion along the EW direction.
However, our long slit optical spectrum along the EW direction which was
analyzed in Paper I shows no hint on the rotational motion;
$\Delta V\lesssim 50$ km s$^{-1}$, suggesting strongly that
the galaxy is seen from an almost face-on view.
Therefore the oval shape of Mrk 1259 may not be due to the
inclination\footnote{It seems no surprise even if an isolated galaxy shows
some morphological peculiarity because any galaxy would experience some
minor merger events in its life. It is also noted that minor mergers can
cause nuclear starbursts (e.g., Hernquist \& Mihos 1995;
Taniguchi \& Wada 1996).}.

\section{OBSERVATIONS}

The $^{12}$CO ($J=1$-0) observations of Mrk 1259 were made using the 45m
radio telescope at Nobeyama Radio Observatory,
equipped with an SIS100 receiver in February 1997.
The observations were made in the midnight under little wind velocity.
Pointing check was made in every 30 minutes and
the pointing errors in our observations are smaller than a few arcsec.
The half power beam width (HPBW) is 15 arcsec at 115 GHz,
corresponding to $\simeq$ 2 kpc at the distance of Mrk 1259.
We observed the radio center position
[$\alpha_{1950}=10^{\rm h}36^{\rm m}02\fs8$;
$\delta_{1950}=-06\arcdeg54\arcmin37\arcsec$; Sramek \& Weedman (1986)]
and four off-nuclear positions (15$^{\prime\prime}$N, 15$^{\prime\prime}$W,
15$^{\prime\prime}$S, and 15$^{\prime\prime}$E).
The main beam efficiency ($\eta_{\rm mb}$) was 0.51 at 115 GHz.
The backend was the 2048 channel wide-band acousto optical spectrometer
operated with the band width of 250 MHz, covering a velocity range of
650 km s$^{-1}$.
The final data were box-car averaged, resulting in a final
velocity resolution of 5 km s$^{-1}$.
The final spectra are shown in Figure 1.

\section{RESULTS AND DISCUSSION}

In Table 1, we give a summary of our observational results.
The integrated CO intensity was estimated by
$I({\rm CO}) = \int T_{\rm A}^* \eta_{\rm mb}^{-1} dv$ K km s$^{-1}$
where $\eta_{\rm mb} = 0.51$.
Using a galactic conversion factor,
$N_{\rm H_{2}}/I_{\rm CO} = 3.6\times 10^{20}$ cm$^{-2}$ (K km s$^{-1}$)$^{-1}$
(Scoville et al. 1987), we estimate the molecular gas mass,
$M_{\rm H_2} = 5.8 \times 10^6 I({\rm CO}) A$, in each position
where $A$ is the projected area of a 15$^{\prime\prime}$ HPBW
in units of kpc$^2$.
For the off-nuclear regions, we also give total values of $I$(CO)
and $M_{\rm H_2}$.
The total molecular gas mass detected in our observations amounts to
$1.2 \times 10^9 M_\odot$.
Since we do not observe the entire disk of this galaxy, this mass
is regarded as a lower limit.

\subsection{The Nuclear CO Emission}

The nuclear CO emission shows a single-peaked profile
with the evident blueward asymmetry.
Applying a two-component Gaussian profile fitting (see the midst panel of
Figure 1), we obtain
the blueshifted broad component with FWHM $\simeq$ 114 km s$^{-1}$
and the narrow one with FWHM $\simeq$ 68 km s$^{-1}$.
The peak velocity of the broad component is blueshifted by 27 km s$^{-1}$
with respect to that of the narrow one (Table 1).
Both the intensities are nearly the same.
We also mention that the red wing cannot be seen in the nuclear CO profile.

Even though there is the broad CO emission component, its width is
significantly narrower than those observed for typical starburst galaxies;
e.g., FWHM(CO) $\simeq$ 200 - 250 km s$^{-1}$
for M82 (Young \& Scoville 1984; Nakai et al. 1987),
$\sim 350$ km s$^{-1}$ for NGC 1808 (Aalto et al. 1994),
and $\sim 325$ km s$^{-1}$ for NGC 4945 (Dahlem et al. 1993).
This difference can be attributed to the effect of viewing angles
between Mrk 1259 and the other starburst galaxies.
It is remembered that the CO line width is generally affected by the galactic
rotation.
Given a typical rotation velocity of a disk galaxy, $V_{\rm rot}
\sim 200$ km s$^{-1}$, the observed full widths would amount to
2$V_{\rm rot} \sim$ 400 km s$^{-1}$ if seen from the edge-on view.
In fact, since we observe M82, NGC 1808, and NGC 4945
from highly inclined viewing angles, their line widths are considered to be
broadened by the effect of galactic rotation.
On the other hand, since Mrk 1259 appears to be a nearly face-on galaxy,
the observed width is not affected by the galactic rotation.
Irwin \& Sofue (1996) suggested that one of the nearby superwind galaxies,
NGC 3628, has a nuclear molecular gas disk with a velocity dispersion
of $\sim$ 100 km s$^{-1}$.
If Mrk 1259 has also such a nuclear gas disk, we can explain the
velocity width of the nuclear CO emission.
Therefore, it is suggested that the observed FWHM of Mrk 1259 is due mainly
to the broadening by some dynamical effect of the starburst activity.
The blueward asymmetry of the nuclear CO line profile suggests that
the CO gas is affected significantly by the superwind.

\subsection{The Off-Nuclear CO Emission}

The detection of the off-nuclear CO emission from Mrk 1259 is very
intriguing from the following two points.
The first point is that the host galaxy of Mrk 1259 appears to be an S0
galaxy (RC3).
It is often observed that early type galaxies such as S0 and
elliptical galaxies tend to have less molecular gas (e.g.,
Young \& Scoville 1991)
although CO emission has been detected from a number of S0 galaxies
(Thronson et al. 1989; Wiklind \& Henkel 1989; Sage 1989; Sage \& Wrobel 1989).
It is also known that the molecular gas in (non-active) S0 galaxies
tends to be concentrated in the region whose diameter is typically
less than one tenth of the optical diameter (Taniguchi et al. 1994).
If this is also the case for Mrk 1259, the molecular gas would be
concentrated within the central $12\arcsec =0.1 D_{\rm 0}$ region
where $D_{\rm 0}$ is the isophotal optical diameter (RC3).
Therefore, the presence of the bright off-nuclear CO emission is one of
very important characteristics of Mrk 1259.

The second point is that the line widths of the off-nuclear CO emission
are comparable to that of the nuclear CO emission, FWHM $\sim 100$ km s$^{-1}$.
If there were an inclined off-nuclear CO disk, we may explain the wide line
width because of the velocity gradient in the disk.
If this is the case, we would observe that the peak velocity
at 15$^{\prime\prime}$E is
significantly different from that at 15$^{\prime\prime}$W.
However, since our observations show that the velocity field of
the off-nuclear regions is almost symmetric,
this possibility is rejected.
The second possibility is that there are spatially extended starburst regions
and a significant amount of molecular gas is associated with them.
However, radio continuum (1.5 GHz and 5 GHz) images show that the starburst
region of Mrk 1259 is concentrated in the central several arcsec region
(R. A. Sramek 1997, private communication).
Therefore, there is no observational evidence for
active star forming regions in the off-nuclear regions.
As described before, we are observing the disk of Mrk 1259
from nearly a face-on view and thus the CO line width would be as
narrow as $\sim$ 10 km s$^{-1}$ if Mrk 1259 were a normal disk galaxy
(Lewis 1984, 1987; Kamphuis \& Sancisi 1993).
If there were an extended molecular gas disk with a velocity dispersion of
$\sim$ 100 km s$^{-1}$ up to a radius of a few kpc, we could explain the wide
line width.
However, the size of the nuclear gas disk in NGC 3628 is much smaller
($\simeq 230$ pc, or $\sim 0.01 D_{\rm 0}$) than the off-nuclear distance of
Mrk 1259 ($\sim 2$ kpc, or $\sim 0.13 D_{\rm 0}$).
Although we cannot rule out the possibility that Mrk 1259 has such a
very extended molecular gas disk with a large velocity dispersion,
we need further detailed molecular-line observations
to confirm this possibility.
The third possibility is that the off-nuclear CO gas is associated with the
superwind (i.e., blown out from the nuclear region).
Since the ionized-gas superwind is extended to $r \sim 3.3$ kpc
(Paper I), this possibility seems to be quite high.
In fact, such extended CO emission is
detected in M82 at the scale of 600 pc (Nakai et al. 1987)
and even at the larger scale ($\sim 2$ kpc; Sofue et al. 1992).
We discuss this possibility in detail in the next section.

\subsection{Biconical Superwind Model for Mrk 1259}

Since the superwind of Mrk 1259 is observed from nearly the pole-on view,
it is interesting to investigate both the velocity field and the
geometry of the superwind.
In order to perform this, we investigate the off-nuclear CO line profile using
a simple biconical outflow model in which the superwind flows toward
the polar directions symmetrically with its apex at the nucleus.
Such a superwind geometry is expected theoretically by hydrodynamical
numerical simulations (Suchkov et al. 1994) and indeed observed in M82
(e.g., Nakai et al. 1987).
In our model, we assume that the molecular gas can only move
along the cone surface.
We assume that the axis of the cone lies along our line of sight.
The full opening angle of the cone ($\theta$) is not well constrained
by the observations because of its nearly face-on viewing angle.
Therefore we take this as a free parameter although Paper I has suggested as
$\theta \lesssim 90\arcdeg$.
A mean tangential velocity of the ionized gas on the sky can be
estimated as $V_{\rm t, ion}\simeq R_{\rm SW}/T_{\rm SW}\simeq (2.3$ kpc$)
/(5.5\times 10^6$ years)$\simeq 410$ km s$^{-1}$ where $R_{\rm SW}$
is the projected radius of the superwind and $T_{\rm SW}$
is the age of the superwind (Paper I).
We note that the outflow velocity of the molecular gas is {\it slower} than
that of the ionized gas because the molecular gas along the cone surface
is {\it dragged} by the ionized gas, rather than directly {\it pushed out}
(Suchkov et al. 1994).
For example, the model A1 of Suchkov et al. (1994) shows that
the velocity of the outflowing dense gas is slower by a factor of
$\sim 5$ than that of the ionized gas at the age of 8.3 Myr.
In fact, comparing the outflow velocity of the ionized gas (Heckathorn 1972)
with that of the molecular gas (Nakai et al. 1987) of M82,
we find that the outflow velocity of the molecular gas is slower
by a factor of $\sim 3$ than that of the ionized-gas.
Thus, the mean tangential velocity of the molecular gas on the sky
can be $V_{\rm t, mol}=V_{\rm t, ion}/\epsilon$
where $\epsilon$ is the decelerating factor ($\epsilon \simeq 3 - 5$).

We examine if the model can explain the observed
CO line profiles in the off-nuclear regions.
No effect of radiative transfer is included in the model calculation.
We assume that the size of the cone is large enough to cover the whole
off-nuclear regions.
For simplicity, we also assume that the velocity field has a power-law form;
i.e., $V(r) \propto r^a$, with a boundary condition of
$V_{\rm t, ion}$ ($r = 2.3$ kpc) = 410 km s$^{-1}$.
The emissivity (strength of the CO emission per a unit area) is also
assumed to have a power-law form; i.e., $I(r) \propto r^b$.
Although the parameters $a$ and $b$ are not well constrained by
the observations, we adopt $a = 1$ and $b = -1$ as representative values
following the trend seen in M82 (Nakai et al. 1987).
We calculate the model for the cases of
$\theta = 60\arcdeg, 90\arcdeg, 120\arcdeg$, and
$150\arcdeg$ and $\epsilon$ = 1, 2, 3, 4, 5, and 6.
To explain the observed FWZI (Full Width at Zero Intensity) of
the off-nuclear CO emission ($\sim 200$ km s$^{-1}$; see Figure 1),
we find that only models with ($\theta = 90\arcdeg$ and
$\epsilon \simeq 5 - 6$) and ($\theta = 120\arcdeg$ and
 $\epsilon \simeq 3 - 4$) are acceptable.
Models with $\theta = 60\arcdeg$ and $150\arcdeg$ cannot reproduce
the line width for any $\epsilon$.
Therefore, we show our results only for the cases
$\theta=90\arcdeg$ and $120\arcdeg$ and $\epsilon$ = 3, 4, 5, and 6
in Figure 2.
Our simple model demonstrates that the CO line has always
a double-peaked profile for any combinations of the parameters.
The red peak corresponds to the recessing cone while the blue one
corresponds to the advancing cone.
On the other hand, our observations show that the off-nuclear CO lines
have smooth profiles around at the systemic velocity.

We discuss why our simple model cannot reproduce the observed off-nuclear
CO profiles.
One possible idea is a ``swirl''-like velocity field which
is often found in the hydrodynamical numerical
simulations (Tomisaka \& Ikeuchi 1988; Suchkov et al. 1994).
If the outflow actually shows such a complex geometry and/or velocity field,
it is expected that some parts of the emission would contribute to
the core emission and can explain the broad and smooth off-nuclear emission.
In order to understand the molecular-gas superwind of Mrk 1259, detailed molecular-line
observations with higher spatial resolution would be helpful.

\vspace{0.5cm}

We would like to thank the staff of Nobeyama Radio Observatory
for their kind support for our observations.
We thank Naomasa Nakai for useful discussion and encouragement and
Takashi Murayama and Shingo Nishiura for kind assistance of the observations.
We also thank R. A. Sramek for kindly providing us his VLA data.
YO was supported by the Grant-in-Aid for JSPS Fellows
by the Ministry of Education, Culture, Sports, and Science.
This work was financially supported in part by Grant-in-Aids for the Scientific
Research (No. 0704405) of the Japanese Ministry of
Education, Culture, Sports, and Science.

%-----------------------------------------------------------------------------
%\clearpage
\newpage

\begin{table}
\caption{Molecular gas properties of Mrk 1259}
\begin{tabular}{llccc}
\tableline
\tableline
 & & & Nucleus & Off-nucleus \\
\tableline
rms noise & $\delta T_{\rm A}^*$ (K) & & 0.018 & $\sim 0.015$ \\
Flux & $I$(CO)$^{a, b}$ (K km s$^{-1}$) & total & $28.0\pm 1.1$ & $43.8\pm 1.9$$^d$ \\
 & & 15\arcsec N & & $4.2\pm 1.0$ \\
 & & 15\arcsec E & & $18.6\pm 1.0$  \\
 & & 15\arcsec S & & $9.5\pm 1.0$ \\
 & & 15\arcsec W & & $11.5\pm 1.0$ \\
Mass & $M_{\rm H_2}^{b, c}$ ($M_\odot$) & & $(4.8\pm 0.2) \times 10^8$ & $(7.5\pm 0.3) \times 10^8$$^d$ \\
Line profile & & & & \\
& FWHM$_{\rm narrow}$ (km s$^{-1}$) & & 68 & \nodata \\
& $V_{\rm narrow}$ (km s$^{-1}$) & & 2178 & \nodata \\
& FWHM$_{\rm broad}$ (km s$^{-1}$) & & 114 & $\sim 100$$^e$ \\
& $V_{\rm broad}$ (km s$^{-1}$) & & 2151 & $\sim 2170$$^e$ \\
& $I$(broad)/$I$(narrow) & & 1.0 & \nodata \\
\tableline
\end{tabular}
\tablenotetext{a}{$I({\rm CO}) = \int T_{\rm A}^* \eta_{\rm mb}^{-1} dv$ K km s$^{-1}$
where $\eta_{\rm mb}$ = 0.51.}
\tablenotetext{b}{Formal one-sigma error indicated.}
\tablenotetext{c}{A Galactic conversion factor, $N_{\rm H_{2}}/I_{\rm CO} = 3.6\times 10^{20}$ cm$^{-2}$ (K km s$^{-1}$)$^{-1}$
(Scoville et al. 1987), is assumed.}
\tablenotetext{d}{Sum of the fluxes of all the four off-nuclear regions.}
\tablenotetext{e}{No profile fitting was made because of both
the lower signal-to-noise ratio and the non-Gaussian profiles.
The values shown here are typical ones for the off-nuclear regions
except 15$^{\prime\prime}$N.}
\end{table}

%------------------------------------------------------------------------------

\newpage

%----------------------------------------------------------------------
%            Figure Caption
%----------------------------------------------------------------------

\figcaption{
The $^{12}$CO($J=1$-0) spectra of Mrk 1259.
The results of the two-component Gaussian fitting of the nuclear
CO emission profile is also shown in the midst panel
with the fitting residuals.
\label{fig1}}

\figcaption{
Simulated off-nuclear CO line profiles based on the simple biconical outflow
model.
The parameters, $\theta$ and $\epsilon$, are described in the text.
\label{fig2}}


\newpage

\begin{references}

\reference{1}{Aalto, S., Booth, R. S., Black, J. H., Koribalski, B., \&
              Wielebinski, R. 1994, \aap, 286, 365}
\reference{1}{Balzano, V. A. 1983, \apj, 268, 602}
\reference{1}{Dahlem, M., Golla, G., Whiteoak, J. B., Wielebinski, R.,
              H\"uttemeister, S., \& Henkel, C. 1993, \aap, 270, 29}
\reference{1}{de Vaucouleurs, G., de Vaucouleurs, A., Corwin, H. G., Jr.,
              Buta, R. J., Paturel, G., \& Fouqu\'e, P. 1991,
              Third Reference Catalogue of Bright Galaxies (Springer-Verlag)}
\reference{1}{Heckathorn, H. M. 1972, \apj, 173, 501}
\reference{1}{Heckman, T. M., Armus, L., \& Miley, G. K. 1990, \apjs, 74, 833}
\reference{1}{Hernquist, L., \& Mihos, J. C. 1995, \apj, 448, 41}
\reference{1}{Irwin, J. A., \& Sofue, Y. 1996, \apj, 464, 738}
\reference{1}{Kamphuis, J., \& Sancisi, R. 1993, \aap, 273, L31}
\reference{1}{Lewis, B. M. 1984, \apj, 285, 453}
\reference{1}{Lewis, B. M. 1987, \apjs, 63, 515}
\reference{1}{Nakai, N., Hayashi, M., Handa, T., Sofue, Y., Hasegawa, T., \&
              Sasaki, M. 1987, \pasj, 39, 685}
\reference{1}{Ohyama, Y., Taniguchi, Y., \& Terlevich, R. 1997, \apj, 480, L9 (Paper I)}
\reference{1}{Sage, L. J. 1989, \apj, 344, 200}
\reference{1}{Sage, L. J., \& Wrobel, J. M. 1989, \apj, 344, 204}
\reference{1}{Scoville, N. Z., Yun, M. S., Sanders, D. B., Clemens, D. P., \&
              Waller, W. H. 1987, \apjs, 63, 821}
\reference{1}{Sofue, Y., Reuter, H. -P., Krause, M., Wielebinski, R., \&
              Nakai, N. 1992, \apj, 395, 126}
\reference{1}{Sramek, R. A., \& Weedman, D. W. 1986, \apj, 302, 640}
\reference{1}{Suchkov, A. A., Balsara, D. S., Heckman, T. M., \& Leitherer, C.
              1994, \apj, 430, 511}
\reference{1}{Taniguchi, Y., Kawara, K., Nishida, M., Tamura, S., \& Nishida, M. T. 1988, \aj, 95, 1378}
\reference{1}{Taniguchi, Y., Murayama, T., Nakai, N., Suzuki, M., \&
              Kameya, O. 1994, \aj, 108, 468}
\reference{1}{Taniguchi, Y., \& Wada, K. 1996, \apj, 469, 581}
\reference{1}{Thronson, H. A., Jr., Tacconi, L., Kenney, J., Greenhouse, M. A.,
              Margulis, M., Tacconi-Garman, L., \& Young, J. S. 1989,
              \apj, 344, 747}
\reference{1}{Tomisaka, K., \& Ikeuchi, S. 1988, \apj, 330, 695}
\reference{1}{Weedman, D. W., Feldman, F. R., Balzano, V. A., Ramsey, L. W.,
              Sramek, R. A., \& Wu, C. -C. 1981, \apj, 248, 105}
\reference{1}{Wiklind, T., \& Henkel, C. 1989, \aap, 225, 1}
\reference{1}{Young, J. S., \& Scoville, N. Z. 1984, \apj, 287, 153}
\reference{1}{Young, J. S., \& Scoville, N. Z. 1991, \araa, 29, 581}

\end{references}
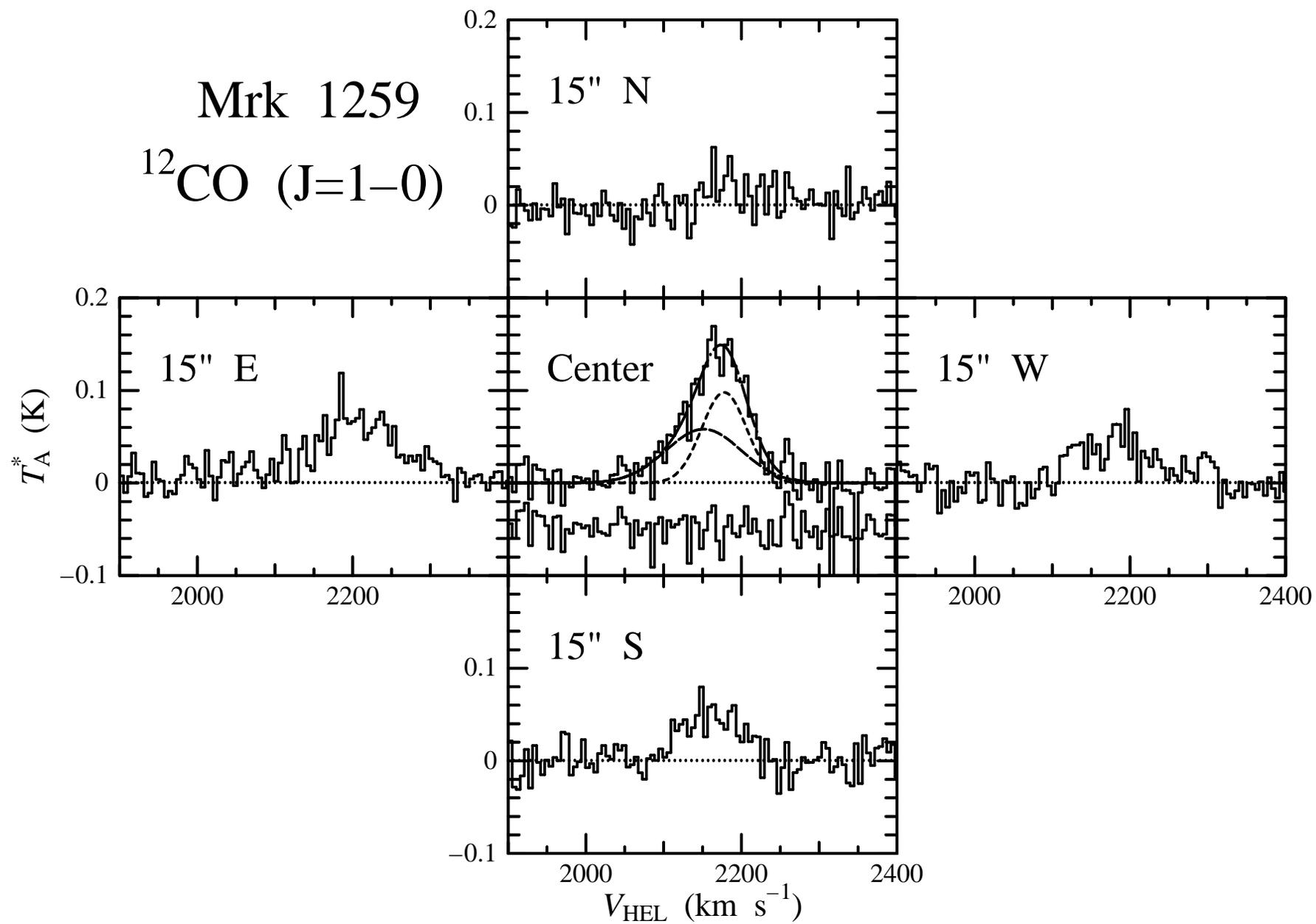
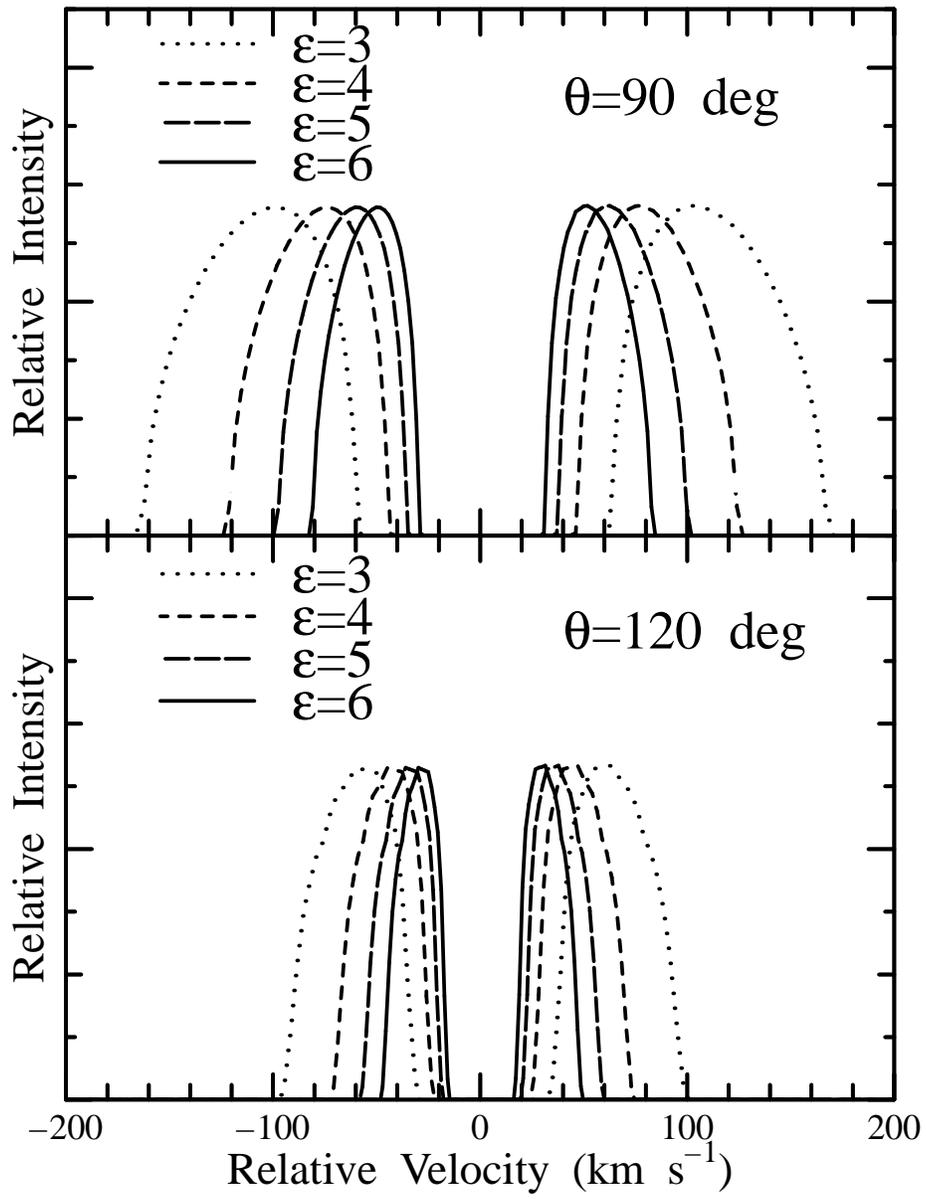
\end{document}